\documentclass{article} 

\usepackage[utf8]{inputenc}

\usepackage{url}
\usepackage{multirow}
\usepackage{lipsum}
\usepackage{amsthm}
\usepackage{mathtools}
\usepackage{xfrac}
\usepackage[export]{adjustbox}
\usepackage{longtable}
\usepackage{booktabs}
\usepackage{colortbl}
\usepackage{xcolor}
\usepackage{graphicx}
\usepackage{subcaption}
\usepackage{verbatim}
\usepackage{placeins}
\usepackage{hyperref}
\hypersetup{
    colorlinks=true,
    urlcolor=[rgb]{0.188, 0.498, 0.886},
    linkcolor=[rgb]{0.188, 0.498, 0.886},    
    filecolor=magenta,      
    citecolor=[rgb]{0.188, 0.498, 0.886},
}
\usepackage{fancyvrb}
\usepackage{fixltx2e}
\usepackage{bera}
\usepackage{pdfpages}

\usepackage{pgf-umlsd}
\usepackage[draft=true]{minted}

\usepackage{algorithm} 
\usepackage{program} 
\usepackage{algorithmic}
\usepackage{listings}
\usepackage{geometry}

\usepackage{caption,setspace}
\DeclareCaptionFormat{listing}{\rule{\dimexpr0.9\columnwidth+17pt\relax}{0.4pt}\par\vskip1pt#1#2#3}
\newcounter{code}



\DeclareMathVersion{sans}
\SetSymbolFont{operators}{sans}{OT1}{cmbr}{m}{n}
\SetSymbolFont{letters}  {sans}{OML}{cmbrm}{m}{it}
\SetSymbolFont{symbols}  {sans}{OMS}{cmbrs}{m}{n}

\lstnewenvironment{sflisting}[1][]
  {\lstset{#1}\mathversion{sans}}{}

\usepackage[normalem]{ulem}

\definecolor{grayalias}{HTML}{3F4444}

\definecolor{bluealias}{HTML}{307FE2}

\title{\texttt{ \textbf{aztarna}}, a footprinting tool for robots}

\author{
    Víctor Mayoral Vilches,
    Gorka Olalde Mendia, 
    Xabier Perez Baskaran,\\ 
    Alejandro Hernández Cordero,
    Lander Usategui San Juan,
    Endika Gil-Uriarte,\\
    Odei Olalde Saez de Urabain and
    Laura Alzola Kirschgens\\
   \small Alias Robotics S.L. \\
   \small Vitoria-Gasteiz, Araba - Álava\\
   \small Spain \\
}

\begin{document}
\includepdf[pages=-, fitpaper]{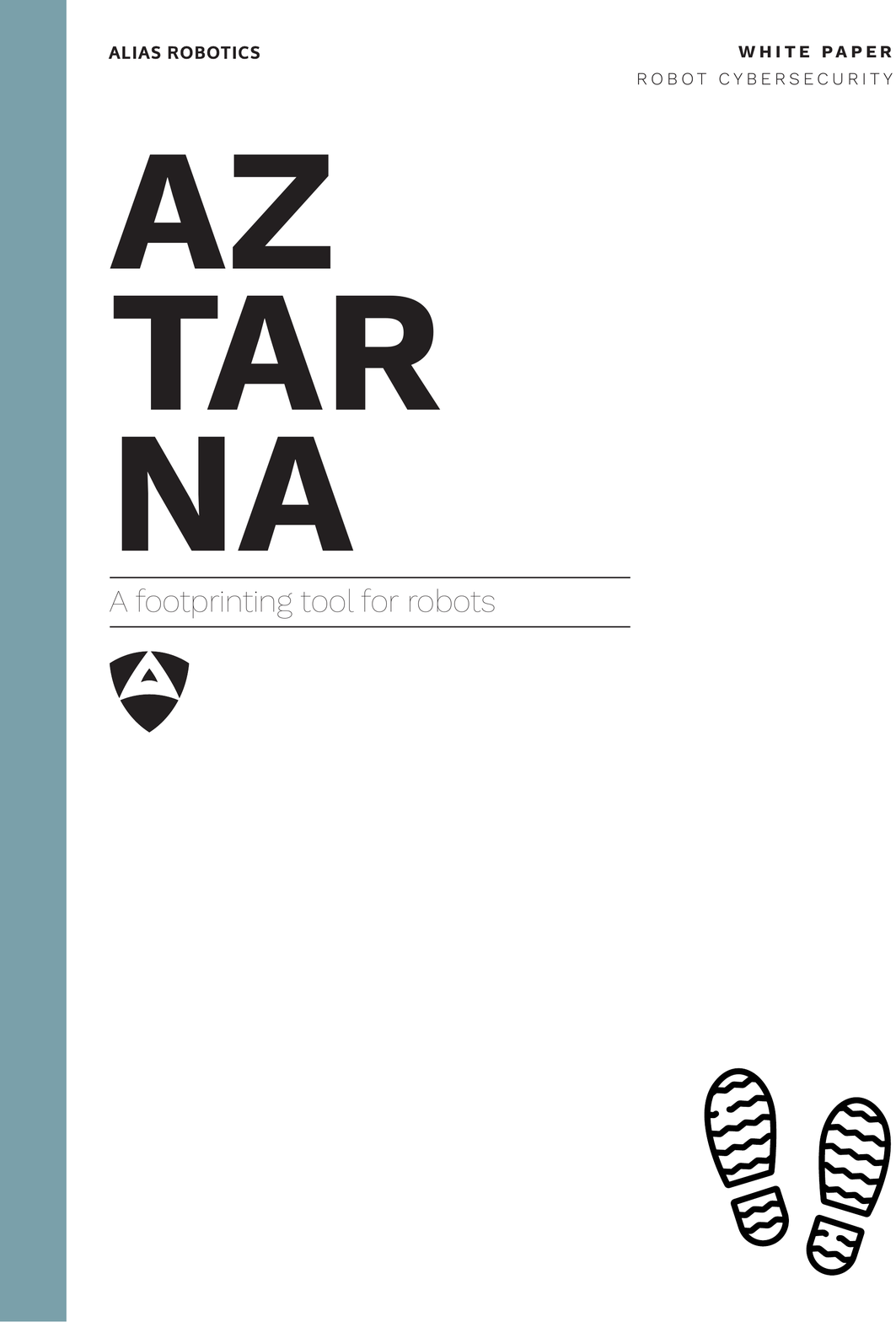}

\maketitle

\vspace{-1em}
\begin{abstract}
Industry 4.0 is changing the commonly held assumption that robots are to be deployed in closed and isolated networks. When analyzed from a security point of view, the global picture is disheartening: robotics industry has not seriously allocated effort to follow good security practices in the robots produced. Instead, most manufacturers keep forwarding the problem to the end-users of these machines. As learned in previous technological revolutions, such as at the dawn of PCs or smartphones, action needs to be taken in time to avoid disastrous consequences. In an attempt to provide the robotics and security communities with the right tools to perform assessments, in this paper we present \texttt{aztarna}, a \emph{footprinting} tool for robotics. We discuss how such tool can facilitate the process of identifying vestiges of different robots, while maintaining an extensible structure aimed for future fingerprinting extensions. With this contribution, we aim to raise awareness and interest of the robotics community, robot manufacturers and robot end-users on the need of starting global actions to embrace security. We open source the tool and disclose preliminary results that demonstrate the current insecurity landscape in industry. We argue that the robotic ecosystem is in need of generating a robot security community, conscious about good practices and empowered by the right tools.

\end{abstract}


\section{Introduction}
Robotics is claimed to be the next technological revolution, and an onset of a new era, dominated by intelligent entities that contribute to human development. Nowadays, robots are no longer only deployed in research oriented organizations, but increasingly handle big amounts of personal and industrial data, perform a variety of automated tasks in industrial scenarios or help humans handle the most hazardous activities. Lately, robot cybersecurity is under question, because recent research shown out the insecurity status of the state of the art in the robotics era \cite{RSF, RVSS} and underlined how downstream implications of vulnerable robots are surpassing those of conventional Information Technologies \cite{RobotHaz}. Some laudable efforts have pointed out an array of internet-exposed robots \cite{DeMarinis2018} that are easily accessible for a remote user by searching for a specific pattern, namely a \emph{robot footprint}. \\

\noindent The example provided by the authors is just a plain particular snapshot of the dooming stage of insecurity of robotics, down to more restrictive scenarios, such as industry, professional environments or simulations and gamification platforms \cite{rctf}. The authors in \cite{DeMarinis2018} mention that the results were surprising for themselves. However, they do not provide additional details or resources to reproduce the method disclosed in the research paper, nor dig more on the nature of the findings and downstream implications.\\ 
\newline
Footprinting, also known as \emph{reconnaissance} is the blueprinting of the security profile of a digital system and its organization, undertaken in a methodological manner. To get this information, typically, a security analyst might use various tools and technologies such as \texttt{whois}, \texttt{nslookup}, \texttt{traceroute}, \texttt{enumerators} or \texttt{pinging}. When applied to robotics, we define robot footprinting as the technique used for gathering information about robots and the entities they belong to. This information becomes very useful when performing security analysis over specific systems. \\

\noindent While footprinting is often understood as a mechanism to obtain network information about a digital system in a generalized manner and using common tools, fingerprinting implies fine tuning the networking requests to elicit a specific signature response from the target device. The procedure allows to obtain additional information such as the Operating System, its version, specific libraries deployed, etc. The boundary between both aspects, footprinting and fingerprinting, is often unclear for new digital systems since one requires the other and depends on the tools available. When looking at robotics, we notice that neither footprinting nor fingerpriting tools have been made available. The direct implication of this fact is that the security researcher, in all cases, needs to develop its own tools.\\
\newline
In an attempt to provide the robotics and security communities with the right tools to perform assessments, we discuss \texttt{aztarna}, a security tool that enables robot footprinting. We discuss how such tool can facilitate the process of identifying blueprints of different robots, while maintaining a extensible structure aimed for future fingerprinting extensions. Section \ref{prior} will introduce some of the prior work and results available. Section \ref{aztarna} will present the \texttt{aztarna} tool, discuss its structure, supported robotic technologies and demonstrate its capabilities through several examples. Section \ref{scan_results} will describe the results obtained while experimenting with \texttt{aztarna}. Finally, section \ref{conclusion} will provide some remarks and share a few pointers towards extensions of \texttt{aztarna} meant for robot fingerprinting.


\section{Previous work}
\label{prior}
\emph{Reconaissance} practice is very common in an increasing number of fields of interest, such as websites, with tactics that contribute to a pre-attack phase. Several \emph{reconaissance} tactiques have arised in the past decades, some of them soundly noticeable, such as massive port scans, whereas others stay unnoticed, below the radar. Internet-wide scans in search for connected machines is a well known practice in the field of cybersecurity, with several first order platforms and tools that help to simplify the process. This is the case of Shodan.io\cite{shodan}, which comprises all the results found by different background scans that are recurrently running in dedicated servers. \\
\newline
Numerous studies have focused upon Industrial Control Systems\cite{Mirian2016} and some other related practices such as Supervisory Control And Data Acquisition (SCADA) control systems. However, not many scans have focused on robots. The work published by the RoboSec project team\cite{Maggi}, with robots of a particular manufacturer in industrial environments, is the positive exception. Yet, the emerging popularity of collaborative robots, which often bring new architectures, as well as the advent of popular robotics frameworks, such as the Robot Operating System (ROS), deserve careful study. At the time of writing, the most pioneering study in the area has been performed by a team of researchers from Brown University \cite{DeMarinis2018}, mainly reporting a proto-footprinting method arising from the findings of ROS instances and robots around the world. The results presented by this team exploited the default configuration of the ROS master, appointed at the 11311 port and available across the Internet. \\
\newline
As introduced above, to the best of our knowledge, the concept of footprinting is diffuse and domain specific in the cybersecurity context. Some researchers claim that the footprinting process is completely passive, by using information publicly available through third party sources. Others empower  both active or passive means of collecting useful information for any target as a reconnaissance step of an attack \cite{McGreevy2001}. In the case of fingerprinting, the existing literature refers to it as "determining the nature of the target by comparing signs provided by the responses of the target against databases or known responses that determine the OS or the application in use". In case of web technologies, user fingerprinting could refer to the techniques of indistinctly determining the user, through a signature or fingerprint generated by his/her browsing activity and via signs left by the browser and OS in use\cite{laperdrix:hal-01285470}. \\

\noindent When applying this to the context of robotics, given that the target scope is not a complete organization, but a specific set of devices in that organization\footnote{Note that we understand robots as cyber-physical systems composed by at least three different elements, sensors to perceive the world, actuators to have a physical impact on it and cognition devices which coordinate the information from sensors and command actuators.}, the meaning of footprinting and fingerprinting changes accordingly. In this context, footprinting could mean obtaining all the possible information from a single robot or group of robots, whereas fingerprinting could be denoted as obtaining information that can unequivocally identify a single robot device, e.g. a serial number, its OS version or particular details about its robotics framework layout and configuration.\\

\noindent The present piece of research aims to improve, systematize and extend the results of previous studies \cite{Maggi, DeMarinis2018} while empowering security researchers with robot footprinting tools. We target ROS as well as other relevant and more secure robot setups, such as Secure ROS (SROS) or ROS 2. Beyond robotics frameworks, our work also targets other robots that do not necessarily employ these popular middlewares. Throughout the following sections, we disclose and describe an open source robot footprinting tool named \texttt{aztarna}, that can be used to reproduce our work and allows for future extensions thanks to its architecture. We discuss such modular architecture and present the initial set-of-supported robot technologies. Ultimately, we discuss against a security by obscurity approach and instead advocate for hacker-tested robot security. While by no means Alias Robotics encourages or promotes unauthorized tampering of running robotic systems. Instead we value the  importance to empower security researchers and raise security-awareness among roboticists, by releasing an robot security auditing tool.



\section{aztarna}
\label{aztarna}

\begin{figure}[htbp]
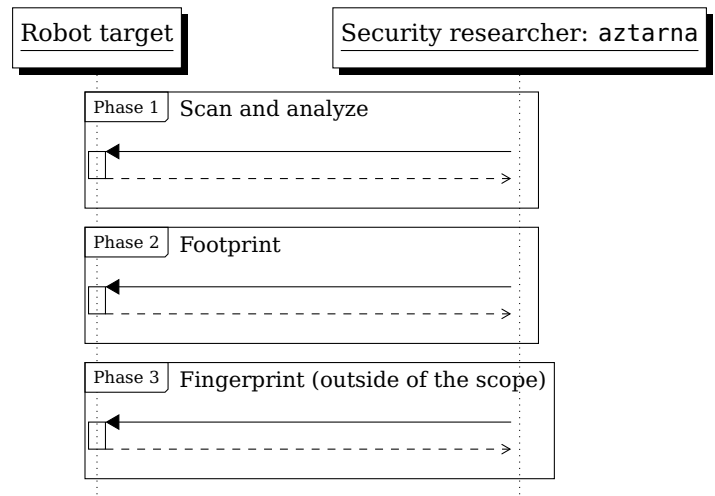

 \centering
 \tikzset{
  every picture/.append style={
    scale=0.2
  }
}
\begin{sequencediagram}
	\renewcommand\unitfactor{1.8}
	\newinst{Robot}{Robot target}
	\newinst[10.0]{Aztarna}{Security researcher: \texttt{aztarna}}
    \begin{scope}[font = \scriptsize]{ }
    	
    	\begin{sdblock}{Phase 1}{Scan and analyze}
    		\begin{call}{Aztarna}{}{Robot}{}
    		\end{call}
	    \end{sdblock}
	
	    \begin{sdblock}{Phase 2} {Footprint}
        	\begin{call}{Aztarna}{}{Robot}{}
        	\end{call}
        \end{sdblock}		    

	    \begin{sdblock}{Phase 3} {Fingerprint (outside of the scope)}
        	\begin{call}{Aztarna}{}{Robot}{}
        	\end{call}

    	\end{sdblock}	
	          
   \end{scope}
\end{sequencediagram}
 \caption{Sequence diagram of \texttt{aztarna}.}
 \label{fig:aztarna/general}
\end{figure}

In Basque language, \texttt{aztarna} means blueprint or mark, in its various forms and meanings. Those include footprint (\emph{aztarna} in Basque) and fingerprint (\emph{hatz-aztarna}). Even in such an ancient language, this same word may be used to name marks upon a given surface and also describe \emph{vestiges appertaining to a concrete organism or entity}. Thus, the pre-romanic language was accurate at the time when defining a word that adheres very well to the  definition of robot footprinting; which basically stands for the information gathering on the \emph{reconnaissance} phase. Thereafter, particular identification deepens into each particular robot through a process that requires crafting specific requests. Such process, known as \emph{fingerprinting} deepens into unique identifiers or features of a particular robot, e.g. OS, libraries, versions or particular communication middlewares signatures.\\
\newline
Motivated by the lack of dedicated tooling for security research in the field of robotics, we have developed \texttt{aztarna}, a tool aimed to help in the detection and scan of robots and robot technologies (including software components) on a network. The tool, developed in Python 3, helps to search for connected robots and gathers some information from those found.\\

\noindent Figure \ref{fig:aztarna/general} illustrates the philosophy behind the \texttt{aztarna} tool. There are three well identified phases that may repeat for each target. For the purpose of this article, our work will focus on the first two phases -scanning and footprinting-, leaving fingerprinting for future work. The architecture of \texttt{aztarna} has been designed to favour its extension towards more and more robotic technologies. The structure of the files within the tool is illustrated in listing \ref{listing:aztarnafiles} where lines 14, 16 and 18  show three folders that contain robot technology-specific code for its footprinting, namely robot adapters. Further extensions can follow a similar pattern and implement the corresponding functions enabling additional robot technology.

\begin{listing}[ht]
\caption{{\texttt{aztarna} code structure, simplified version.}}
\begin{minted}
[
frame=lines,
framesep=2mm,
baselinestretch=1.2,
fontsize=\footnotesize,
linenos=true
]{bash}
aztarna/
    Dockerfile
    README.md
    aztarna
        init__.py
        __main__.py
        cmd.py
        commons.py
        helpers.py
        ros
            init__.py
            commons.py
            helpers.py
            ros
              ...
            sros
              ...
        industrialrouters
            helpers.py
            scanner.py
              ...
    docs
        Makefile
        ...
...
\end{minted}    
\label{listing:aztarnafiles}
\end{listing}

\noindent \texttt{aztarna} has different work modes that allow to use the tool in different scenarios, and together with other tools. The robot footprinting tool provides flexibility when deciding on the hosts to scan, which can be loaded from an input file, determined by a network IP range, or even loaded from \texttt{stdin} as part of a pipe. This allows to use \texttt{aztarna} in conjunction with tools aimed for massive scans such as \cite{Durumeric2013} ZMap, to scan vast amounts of hosts, even the whole Internet network range. Regarding the ports to scan, the tool also allows to choose between a single port, a range of ports or a port list.\\

\noindent For large scan performance improvement, \texttt{aztarna} provides a basic and a extended mode of scan. With the same purpose, extensive usage of asynchronous development has been used, with the help of Python AsyncIO. This allows the application to handle a big number of concurrent connections without the usage of threads, and improves the performance substantially in comparison to them.\\

\noindent The results provided by \texttt{aztarna} can be exported to standard CSV files, containing all the data gathered from the nodes. This allows to employ results for future analysis. The output file contains a common structure including all the findings, that when exported to third party tools provide ways to filter the results by resource type, names, addresses, and therefore, by found robots. The usage of \texttt{aztarna} is straightforward, as all the different behaviours are defined by command line parameters, which are described when calling the tool with no parameters. \\

\noindent In the following section we will briefly cover \texttt{aztarna}'s robot adapters, the abstraction used to support additional robot technologies. We will discuss the structure of a robot adapter and introduce a few examples.


\subsection{Robot adapters}
\label{robadapt}

Similar to what happened in the computer industry, there is a plurality of robot manufacturers, each using its own hardware and software. As a tool to footprint robots and robot-related technology, \texttt{aztarna} provides a common skeleton that can easily be extended to support new software or hardware robot components. New components are extended via \emph{robot adapters}, abstractions that enlarge the base class \texttt{RobotAdapter}. Robot adapters provide the methods to footprint the corresponding robot technology and are typically organized in folders.\\
\newline
The sections below describe some of the supported robots and robot components within \texttt{aztarna}.

\subsubsection{Robot Operating System (ROS) adapter}

In the case of ROS, the connection is made directly to the master via XMLRPC, from which all information sent by nodes is inferred. This information consists on nodes, topics, services, parameters and all the interactions between them, including subscriptions and publications.


\begin{figure}[htbp]
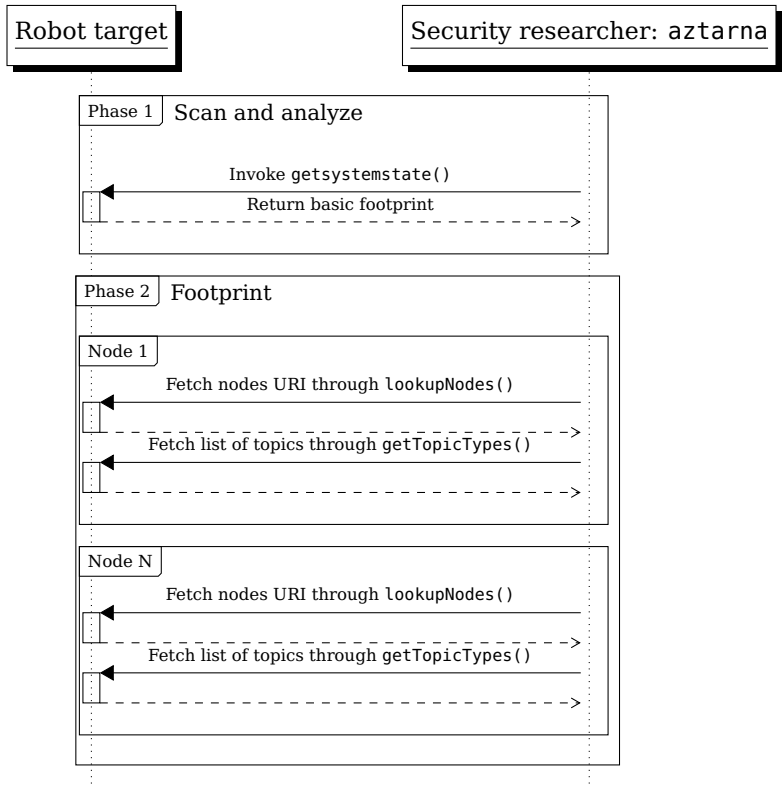

 \centering
 \tikzset{
  every picture/.append style={
    scale=0.2
  }
}
\begin{sequencediagram}
	\renewcommand\unitfactor{2.0}
	\newinst{Robot}{Robot target}
	\newinst[15.0]{Aztarna}{Security researcher: \texttt{aztarna}}
    \begin{scope}[font = \scriptsize]{ }
    	\begin{sdblock}{Phase 1}{Scan and analyze}
    	\stepcounter{seqlevel}
		\begin{call}{Aztarna}{Invoke \texttt{getsystemstate()}}{Robot}{Return basic footprint}
		\end{call}
		
	\end{sdblock}
	\begin{sdblock}{Phase 2} {Footprint}	

	\begin{sdblock}{Node 1} {}
    	\begin{call}{Aztarna}{Fetch nodes URI through     \texttt{lookupNodes()}}{Robot}{}
    	\end{call}
        
        \begin{call}{Aztarna}{Fetch list of topics through \texttt{getTopicTypes()}}{Robot}{}
    	\end{call}
    	
	\end{sdblock}

	\begin{sdblock}{Node N} {}
    	\begin{call}{Aztarna}{Fetch nodes URI through     \texttt{lookupNodes()}}{Robot}{}
    	\end{call}
        
        \begin{call}{Aztarna}{Fetch list of topics through \texttt{getTopicTypes()}}{Robot}{}
    	\end{call}
    	
	\end{sdblock}	
         
	          
   \end{sdblock}		
   \end{scope}
\end{sequencediagram}
 \caption{Sequence diagram of a \texttt{aztarna} scanning a ROS target.}
 \label{fig:aztarna/ros}
\end{figure}

\noindent Figure \ref{fig:aztarna/ros} provides insight on the Phases 1 and 2 of \texttt{aztarna} for ROS\footnote{Logic implemented and available at \url{https://github.com/aliasrobotics/aztarna/tree/master/aztarna/ros/ros}}.\\

\noindent The usage of the \texttt{aztarna}'s ROS robot adapter is demonstrated in Listing \ref{listing:robotadapter_ros}. The tool is invoked with the \texttt{-t ROS} flag, indicating that the robot adapter should use ROS one. Furthermore, flags \texttt{-p} and \texttt{-a} mean the range of ports and addresses to be scanned.

\begin{listing}[ht]
\caption{{\texttt{aztarna} using the ROS robot adapter.}}
\begin{minted}
[
frame=lines,
framesep=2mm,
baselinestretch=1.2,
fontsize=\footnotesize,
linenos
]{bash}
root@3c22d4bbf4e1:/# aztarna -t ROS -p 11311-11320 -a 127.0.0.1
root@432b0c5f61cc:~/aztarna# aztarna -t ROS -p 11311-11320 -a 127.0.0.1
[-] Error connecting to host Address: 127.0.0.1: Cannot connect to host 127.0.0.1:11315 ssl:None [Connection refused]
	Not a ROS host
[-] Error connecting to host Address: 127.0.0.1: Cannot connect to host 127.0.0.1:11312 ssl:None [Connection refused]
	Not a ROS host
...
[+] ROS Host found at 127.0.0.1:11317
[+] ROS Host found at 127.0.0.1:11311
\end{minted}    
\label{listing:robotadapter_ros}
\end{listing}

\noindent Listing \ref{listing:robotadapter_ros} mainly represents Figure's \ref{fig:aztarna/ros} Phase 1: \emph{Scan and analyze}. Making use of the ROS Master API, \texttt{aztarna} sends \texttt{getsystemstate()} requests to the corresponding ports and addresses. Based on the responses received, the tool analyses the information and determines which host-port combination contains a ROS Master.\\
\newline
\texttt{aztarna} can also be launched with additional flags that allow the tool to perform footprinting actions. In particular and as depicted in Figure \ref{fig:aztarna/ros} Phase 2: \emph{Footprint}, when using the \texttt{aztarna}'s flag \texttt{-e}, more information about a particular ROS host can be obtained through the exploitation of the ROS Master API. Listing \ref{listing:robotadapter_ros_footpriting} shows en example.\\

\begin{listing}[ht]
\caption{{\texttt{aztarna} using ROS robot adapter to perform footprinting (Phase 2 from Figure \ref{fig:aztarna/ros})}}
\begin{minted}
[
frame=lines,
framesep=2mm,
baselinestretch=1.2,
fontsize=\footnotesize,
linenos
]{bash}
root@aa6b6d7f9bd3:/# aztarna -t ROS -p 11311 -a 127.0.0.1 -e
[+] ROS Host found at 127.0.0.1:11311

Node: /rosout XMLRPCUri: http://aa6b6d7f9bd3:39719

	 Published topics:
		 * /rosout_agg(Type: rosgraph_msgs/Log)

	 Subscribed topics:
		 * /rosout(Type: rosgraph_msgs/Log)

	 Services:
		 * /rosout/set_logger_level
		 * /rosout/get_loggers

	 CommunicationROS 0:
		 - Publishers:
		 - Topic: /rosout(Type: rosgraph_msgs/Log)
		 - Subscribers:
			/rosout XMLRPCUri: http://aa6b6d7f9bd3:39719

	 CommunicationROS 1:
		 - Publishers:
			/rosout XMLRPCUri: http://aa6b6d7f9bd3:39719
		 - Topic: /rosout_agg(Type: rosgraph_msgs/Log)
		 - Subscribers:
\end{minted}    
\label{listing:robotadapter_ros_footpriting}
\end{listing}

\noindent When combined with the ROS robot adapter, \texttt{aztarna} provides security researchers the tools to inspect and footprint ROS deployments across different networks. Moreover, the tool can be conjoint with other Linux commands to perform complete analysis over particular systems. Listing \ref{listing:robotadapter_ros_footpriting_all} provides an example of how \texttt{aztarna} can be used to find unmodified ROS instances in particular machines using its loopback virtual network interface.

\begin{listing}[ht]
\caption{{\texttt{aztarna}'s use case which checks for all ROS instances in the loopback virtual interface of a given machine}}
\begin{minted}
[
frame=lines,
framesep=2mm,
baselinestretch=1.2,
fontsize=\footnotesize,
linenos
]{bash}
root@bc6af321d62e:/# nmap -p 1-65535 127.0.0.1 | grep open | awk '{print $1}' | sed "s*/tcp**" | \ 
    sed "s/^/aztarna -t ROS -p /" | sed "s/$/ -a 127.0.0.1/" | bash
[+] ROS Host found at 127.0.0.1:11311
[+] ROS Host found at 127.0.0.1:11317
[-] Error connecting to host 127.0.0.1:38069 -> Unknown error
	Not a ROS host
[-] Error connecting to host 127.0.0.1:38793 -> Unknown error
	Not a ROS host
[-] Error connecting to host 127.0.0.1:45665 -> <type 'exceptions.Exception'>:method "getSystemState" is not supported
	Not a ROS host
[-] Error connecting to host 127.0.0.1:46499 -> <type 'exceptions.Exception'>:method "getSystemState" is not supported
	Not a ROS host
...
\end{minted}    
\label{listing:robotadapter_ros_footpriting_all}
\end{listing}

\FloatBarrier

\subsubsection{Secure ROS (SROS) adapter}

\begin{figure}[htbp]
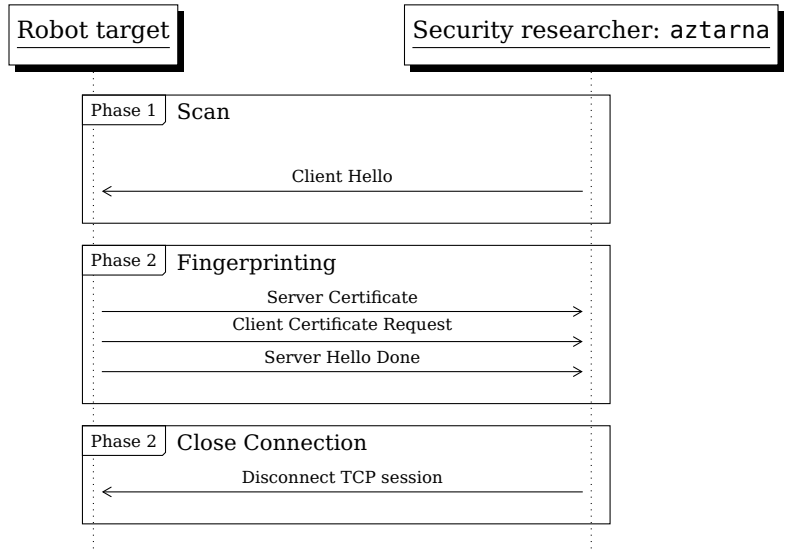

 \centering
 \tikzset{
  every picture/.append style={
    scale=0.2
  }
}
\begin{sequencediagram}
	\renewcommand\unitfactor{2.0}
	\newinst{Robot}{Robot target}
	\newinst[15.0]{Aztarna}{Security researcher: \texttt{aztarna}}
    \begin{scope}[font = \scriptsize]{ }
    	\begin{sdblock}{Phase 1}{Scan}
    	\stepcounter{seqlevel}
		\mess{Aztarna}{Client Hello}{Robot}{Server Hello}
	\end{sdblock}
	\begin{sdblock}{Phase 2}{Fingerprinting}
		\mess{Robot}{Server Certificate}{Aztarna}
		\mess{Robot}{Client Certificate Request}{Aztarna}
		\mess{Robot}{Server Hello Done}{Aztarna}
	\end{sdblock}
	\begin{sdblock}{Phase 2}{Close Connection}
		\mess{Aztarna}{Disconnect TCP session}{Robot}
	\end{sdblock}
   \end{scope}
\end{sequencediagram}
 \caption{Sequence diagram of a \texttt{aztarna} performing the scan and fingerprinting phases for SROS.}
 \label{fig:aztarna/sros}
\end{figure}

\noindent In the case of SROS, as the connection to the master is not possible due to the requirement of a client certificate, the policies that are publicly available in the certificate presented by the server are used\cite{White} to gather information. These policies come in the form of standard x509 certificate policies and consist of an unique object identifier(OID), as well as in an optional classifier\cite{rfc5280}, being able to introduce wildcard characters in order to cover a group of target objects.\\

\begin{listing}[ht]
\caption{{Certificate policies present in \textit{rosout.py} node certificate}}
\begin{minted}
[
frame=lines,
framesep=2mm,
baselinestretch=1.2,
fontsize=\footnotesize,
linenos
]{bash}
            X509v3 Certificate Policies: critical
                Policy: 1.2.3.4.5.6.7.8.9.1.1
                  CPS: /rosout
                Policy: 1.2.3.4.5.6.7.8.9.2.1
                  CPS: /rosout
                  CPS: /rosout_agg
                Policy: 1.2.3.4.5.6.7.8.9.3.2
                  CPS: **
                Policy: 1.2.3.4.5.6.7.8.9.4.1
                  CPS: /rosoutpy/get_loggers
                  CPS: /rosoutpy/set_logger_level
                Policy: 1.2.3.4.5.6.7.8.9.5.1
                  CPS: /enable_statistics
                  CPS: /tcp_keepalive
                  CPS: /use_sim_time
                Policy: 1.2.3.4.5.6.7.8.9.6.2
                  CPS: **

\end{minted}    
\label{listing:robotadapter_sros_policies}
\end{listing}

\noindent Policies on SROS are defined to specify which topics a node can subscribe or publish to, which services it can call, as well as which parameters can it read or write.\\ 
Each of this policies consist of a unique OID for each of the nodes and types of policies, as well as multiple values for each of the policies. An example of the policies for the \textit{rosout.py} node is shown in Listing \ref{listing:robotadapter_sros_policies}.\\

\noindent As the standard libraries available for managing TLS connections in Python do not support obtaining the server certificate in those cases where the connection is not completed, manual handling of the connection is required. For that purpose, Scapy\cite{biondi} library is used, along with the TLS support layer available in the library.\\

\noindent In the case of the SROS adapter, it is no possible to infer the information only from the master, as the policies are unique for each node certificate.
To obtain all the information about the system, a port scan on the host is required to determine the active nodes.\\

\noindent For each of these nodes, a TLS connection is performed by which the policies are obtained. The policies for each of the nodes show the resources that the given node has allowed or denied the access to. These resources include publication or subscription to topics, calls to services and read or modification of parameters.
Due to the extended timespan required to scan SROS hosts, the \texttt{-e} flag has special noteworthiness. Launching a plain scan will result only in scanning the master node, whereas launching the scan with the \texttt{-e} flag will cause to perform the port scan and connection to each of the found ports.\\

\noindent Apart from the information provided by the certificate policies, valuable information is also acquired from the certificate subject fields. It is a common practice for organizations to set valid information regarding to the organization itself, as well as for the person or the organization in charge of issuing the certificates. This data is considered very valuable in reconnaissance scenarios. In the field of interest, this subject could identify the owner corporation of the found robot, as well as the person responsible for the secure configuration of the robots, which could ease the effort required to determine a possible target for social-engineering attack in the case of a malicious actor.\\

\noindent Listing \ref{listing:robotadapter_sros_certificate_info} shows an excerpt of the output showing the subject and issuer information for the certificate chain of the Google website.
\begin{listing}[ht]
\caption{{OpenSSL call fetching Google website certificate issuer information.}}
\begin{minted}
[
frame=lines,
framesep=2mm,
baselinestretch=1.2,
fontsize=\footnotesize,
linenos
]{bash}
bash-3.2$ openssl s_client -connect www.google.com:443
CONNECTED(00000008)
depth=2 OU = GlobalSign Root CA - R2, O = GlobalSign, CN = GlobalSign
verify return:1
depth=1 C = US, O = Google Trust Services, CN = Google Internet Authority G3
verify return:1
depth=0 C = US, ST = California, L = Mountain View, O = Google LLC, CN = www.google.com
verify return:1
\end{minted}    
\label{listing:robotadapter_sros_certificate_info}
\end{listing}

\noindent In the case of SROS, apart of fetching the subject information, a check is performed in order to determine if the demo setup is being used.\\
The certificates issued by the demo setup provided by SROS always contain the same values as the issuer, which causes them to be easily recognizable, specially given that the setup includes a typo on the State field, calling it \textit{Sate}, as shown in Listing  \ref{listing:robotadapter_sros_certificate_info_demo}

\begin{listing}[H]
\caption{{OpenSSL call showing SROS demo setup certificate.}}
\begin{minted}
[
frame=lines,
framesep=2mm,
baselinestretch=1.2,
fontsize=\footnotesize,
linenos
]{bash}
bash-3.2$ openssl s_client -connect 127.0.0.1:11311
CONNECTED(00000006)
depth=2 ST = Sate, O = Organization, C = ZZ, OU = Organizational Unit, CN = root, L = Locality
verify error:num=19:self signed certificate in certificate chain
verify return:0
\end{minted}    
\label{listing:robotadapter_sros_certificate_info_demo}
\end{listing}

\noindent While not a issue by itself, the usage of the demo setup could give an insight on the security posture of a robot user. It could also be a good metric to define if the system is for testing purposes or in a real production environment.\\

\noindent In order to perform a scan with aztarna in search for SROS based systems, the \texttt{-t SROS} argument is used. This type of scan searches for a master running SROS, establishes a connection for collecting the server certificate and finally recovers the data from this certificate. In this case, given that the master doesn't have any policy information, only the information regarding to the certificate subject is gathered and the check for the demo setup is performed. In Listing \ref{listing:robotadapter_sros_scan_simple}, the output of an example scan is shown.

\begin{listing}[h!]
\caption{{\texttt{aztarna} using SROS robot adapter to perform a basic footprinting scan.}}
\begin{minted}
[frame=lines,
framesep=2mm,
baselinestretch=1.2,
fontsize=\footnotesize,
linenos
]{bash}
bash-3.2$ ./aztarna -t SROS -a 192.168.64.131
Connecting to 192.168.64.131:11311
[+] SROS host found!!!
192.168.64.131:11311
        Node name: master
        Port: 11311
        Demo CA Used: True
\end{minted}    
\label{listing:robotadapter_sros_scan_simple}
\end{listing}

\noindent For the purpose of gathering information regarding to all the nodes in the system, an extended scan must be performed. The extended scan comprises a full scan of the target host, in which all the ports are checked for existing nodes.\\

\noindent For performing the scan, in the first place, the presence of a SROS master is checked in the selected ports, or in the default 11311 port, if not specified. If a SROS master is found, the tool performs a full port scan on the target host seeking the presence of nodes. For each found port, the tool attempts to establish a TLS connection and gather server certificates. In the same manner, for each of the successfully collected certificates, the node policies and subject information are collected.
Due to the time required and the noise generated during the performance of a full port scan, this scan is only recommended for a low number of target hosts. A simplified example with the output of one extended scan is shown in Listing \ref{listing:robotadapter_sros_scan_extended}.

\begin{listing}[H]
\caption{{\texttt{aztarna} using SROS robot adapter to perform an extended footprinting scan.}}
\begin{minted}
[
frame=lines,
framesep=2mm,
baselinestretch=1.2,
fontsize=\footnotesize,
linenos
]{bash}
bash-3.2$ ./aztarna -t SROS -a 192.168.64.131 -e
Connecting to 192.168.64.131:11311
[+] SROS host found!!!
Scanning host 192.168.64.131:39189
Scanning host 192.168.64.131:35383
...
Scanning host 192.168.64.131:38429
Scanning host 192.168.64.131:11310
...
(IPv4Address('192.168.64.131'), 41369, None)
(IPv4Address('192.168.64.131'), 11310, [X.509 Cert. Subject:/C=ZZ/ST=Sate/L=Locality/
O=Organization/OU=Organizational Unit/CN=keyserver,
Issuer:/C=ZZ/ST=Sate/L=Locality/O=Organization/OU=Organizational Unit/CN=master])
...
192.168.64.131:11311
	Node name: master
	Port: 11311
	Demo CA Used: True

	Node name: talker
	Port: 
	Demo CA Used: True
	Policies:
		Type: Subscriptable topics
		Permission: False
		Values: 
			b'**'

		Type: Publishable topics
		Permission: False
		Values: 
			b'/chatter'
			b'/rosout'

		Type: Unknown
		Permission: False
		Values: 
			b'**'

		Type: Executable services
		Permission: False
		Values: 
			b'/talker/get_loggers'
			b'/talker/set_logger_level'

		Type: Readable parameters
		Permission: False
		Values: 
			b'/use_sim_time'

...
\end{minted}    
\label{listing:robotadapter_sros_scan_extended}
\end{listing}

\subsubsection{Industrial routers}
\label{subsec:irouters}

\begin{figure}[htbp]
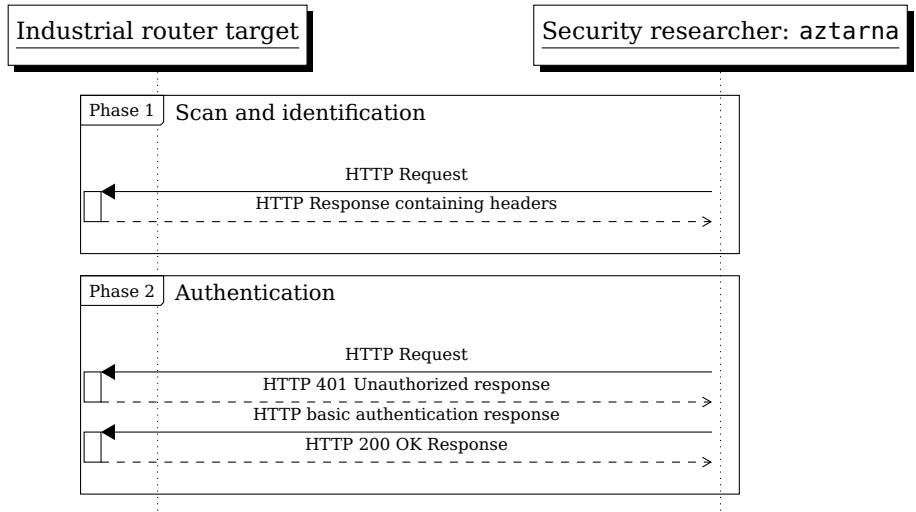

 \centering
 \tikzset{
  every picture/.append style={
    scale=0.2
  }
}
\begin{sequencediagram}
	\renewcommand\unitfactor{2.0}
	\newinst{Router}{Industrial router target}
	\newinst[15.0]{Aztarna}{Security researcher: \texttt{aztarna}}
    \begin{scope}[font = \scriptsize]{ }
    	\begin{sdblock}{Phase 1}{Scan and identification}
        	\stepcounter{seqlevel}
        	\begin{call}{Aztarna}{HTTP Request}{Robot}{HTTP Response containing headers}
        	\end{call}
	    \end{sdblock}
	    \begin{sdblock}{Phase 2}{Authentication}
	        \stepcounter{seqlevel}
	        \begin{call}
	        {Aztarna}{HTTP Request}{Robot}{HTTP 401 Unauthorized response}
	        \end{call}
	        \begin{call}
	        {Aztarna}{HTTP basic authentication response}{Robot}{HTTP 200 OK Response}
	        \end{call}

	    \end{sdblock}
         
	          
   \end{scope}
\end{sequencediagram}
 \label{fig:aztarna/industrial_routers}
 \caption{\texttt{aztarna} performing identification and authentication phases for an industrial router.}
\end{figure}

\noindent Direct Internet exposure may seem unrealistic in industrial environments where an external attacker can tamper critical configuration data and modify the behaviour of robots. Yet, there exist cases where industrial routers are reachable from outside their operating network. Some of them use default credentials or even worse, unrestricted access.\\

\noindent There is a new trend in industrial robots where a connection is opened to the open internet so as to get, to name a few, over-the-air updates, maintenance or monitoring. These routers allow users to connect to the robot as if you were on a local network. The number of robots connected to the internet is increasing, fact that exposes robots to cyber attacks. In principle, industrial robots were designed to be isolated, but The Internet of the things (IoT) and the Industrial Internet of the Things (IIoT) have evolved to give internet access to robots, making this system a potential target for cyber attackers.\\

\noindent Following the programming approach explained in section \ref{robadapt} a \texttt{Robot Adapter} has been implemented for industrial routers. The selected router brands are Westermo, Moxa, Sierra Wireless and eWON. As commonly the web consoles for this routers reside on widely used ports, such as standard 80 and 443 ports, the idea of scanning the whole internet in search of these turns unbearable. Taking this argument into account, two different strategies are taken, one based on local network host sweeping and the other based on the whole internet scanning (Taking advantage of the Shodan API \cite{shodan}).

\subparagraph{Local Network footprinting} This approach is meant for scenarios where the \texttt{aztarna} tool is run in local networks. In this scenario, the range of available hosts is reduced, meaning that scans on the cited ports are feasible. The usage of aztarna’s industrial robot adapter is demonstrated in Listing \ref{listing:robotadapter_irouter_scan}. The tool is run indicating it should search for web services on ports 80 and 5001 respectively. 

\begin{listing}[H]
\caption{{\texttt{aztarna} using IROUTERS robot adapter to perform a footprinting scan.}}
\begin{minted}
[
frame=lines,
framesep=2mm,
baselinestretch=1.2,
fontsize=\footnotesize,
linenos
]{bash}
$ aztarna -t IROUTERS -a 192.168.1.0/24 -p 80,5001
[+] eWON router in http://192.168.1.10:80 is not secure
[+] Westermo router in http://192.168.1.11:5001 is secure
\end{minted}    
\label{listing:robotadapter_irouter_scan}
\end{listing}

\noindent When possible industrial routers are found, the results are split into two groups: \textit{Secure} and \textit{Not Secure}. \textit{Secure} means that the router is not accessible with default credentials and \textit{Not Secure} means that the router is accessible due to default credential usage or even an unrestricted open access. The technique used for detecting router types consists of grabbing and analyzing the HTTP headers present in the router's response message:

\begin{itemize}
    \item \textbf{Westermo}: These routers responses contains a field called \textit{WWW-Authenticate} which describes the router's model \textit{Westermo ADSL-350}.
\begin{Verbatim}[commandchars=\\\{\},codes={\catcode`$=3\catcode`_=8}]
    HTTP/1.1 401 Unauthorized
    Server: GoAhead-Webs
    Date: Mon Nov 26 01:07:08 2018
     WWW-Authenticate: Basic realm="\textbf{Westermo} ADSL-350"
    Pragma: no-cache
    Cache-Control: no-cache
    Content-Type: text/html
\end{Verbatim}
\item \textbf{eWON}: These routers \textit{Server} response header field directly indicates the router type::

\begin{Verbatim}[commandchars=\\\{\},codes={\catcode`$=3\catcode`_=8}]
    HTTP/1.1 302 Redirect
    Access-Control-Allow-Origin: *
    Access-Control-Allow-Credentials: true
    Server: \textbf{eWON}
    Date: Sun Nov 25 21:05:53 2018 GMT
    Connection: close
    Pragma: no-cache
    Cache-Control: no-cache,max-age=0,must-revalidate
    Content-Type: text/html
\end{Verbatim}

\item \textbf{Moxa}: In the case of Moxa routers, two well differentiated versions exist, each one with different interface and server versions. Old Moxa Routers return a Server header detailing the \textit{MoxaHttp/1.0} versions, while newer ones return a server field with a value of \textit{MoxaHttp/2.2}. This two versions differ in the way that the authentication is performed, which will be described on later sections.

\begin{Verbatim}[commandchars=\\\{\},codes={\catcode`$=3\catcode`_=8}]
    HTTP/1.1 200 OK
    Date: Wed, 19 Feb 2003 09:00:00 GMT
    Server: \textbf{MoxaHttp/1.0}
    Pragma: no-cache
    Cache-Control: no-cache
    Content-type: text/html
    Content-length: 34273
\end{Verbatim}

\begin{Verbatim}[commandchars=\\\{\},codes={\catcode`$=3\catcode`_=8}]
    HTTP/1.1 200 OK
    Date: Mon, 03 Mar 2008 08:00:00 GMT 
    Server: \textbf{MoxaHttp/2.2}
    Pragma: no-cache
    Cache-Control: no-cache
    Content-Type: text/html
    Connection: close
    Transfer-Encoding: chunked
\end{Verbatim}

\item \textbf{Sierra Wireless}: Sierra Wireless routers expose an identifying Server field in the header as well, independently from the router version.

\begin{Verbatim}[commandchars=\\\{\},codes={\catcode`$=3\catcode`_=8}]
    HTTP/1.1 200 OK
    Date: Mon, 03 Mar 2008 08:00:00 GMT 
    Server: \textbf{MoxaHttp/2.2}
    Pragma: no-cache
    Cache-Control: no-cache
    Content-Type: text/html
    Connection: close
    Transfer-Encoding: chunked
\end{Verbatim}

\end{itemize}

\subparagraph{Shodan based footprinting} Contrary to ROS machines, which don't run in well known ports, the connection to industrial routers is usually done in widely used HTTP(80) and HTTPS(443) ports. In this scenario, a manual internet scan for the given ports is impractical, due to the high number of results and false positives obtained. To overcome this issue, the Shodan API has been employed. Shodan is a internet wide scanning platform, in which different crawlers continuously index and store found devices on the internet\cite{shodan}. These devices can then be retrieved by the use of queries. In this case, part of the work is done by \texttt{aztarna}. The detection of each of the router models based on the HTTP request headers, is leveraged to the Shodan API. This scan returns an already filtered set of valid router devices which the tool is able to handle.\\

\noindent To run a full evaluation for each of the industrial router models, \texttt{aztarna} performs a set of actions to obtain the data. First and as mentioned in the paragraph above, the tool employs Shodan to perform the first filtering of targets and retrieve their addresses and ports. Once the target acquisition is done, for each of the industrial routers, a default credentials check is performed. This check employs a set of default credentials harvested from manufacturer documentation and other sources. For each of the credentials, a log-in attempt is made, and successful attempts registered. Finally, additional metadata for the scanned address is obtained with the mediation of the Whois database. For each of the addresses, the country and ASN description are stored.

\subparagraph{Password checking} for each of the router models is considered one of the vital steps in the assessment. The metric of how much routers have default credentials or no authentication at all gives a clear insight on the security posture of the industry against menaces arising from this internet connected devices. 
For evaluating if a router is using default credentials, different login attempts are made to each of the router with passwords that have been documented by the manufacturer or by other sources\cite{scadastrangelove_2016} being used as defaults.\\

\noindent Each router model follows a different schema for authentication, being basic HTTP authentication the most used by manufacturers. This type of authentication is the easiest to check among the tested ones, as the HTTP response code reveals if the login attempt has been successful or not, a 200(OK) code meaning a successful login attempt, and a 401(Unauthorized) code showing a failed login attempt, as shown on the example workflow figure \ref{fig:aztarna/industrial_routers}. Manufacturers not using basic authentication implement various types of HTML forms for user login, some of them including certain types of challenges and client side hashing of passwords.

\begin{itemize}
    \item \textbf{Westermo Routers} present a basic HTTP authentication scheme. The router does not provide any unauthenticated page, so the base url (\texttt{/}) is adequate for checking the credentials.
    
    \item \textbf{Ewon Routers} present a basic HTTP authentication scheme. In contrast with the Westermo routers, this type of routers do allow unauthenticated access to certain status pages. As the pages requiring authentication vary depending of the router model, a common path has been required to be determined. On empirical research, the \texttt{/Ast/MainAst.shtm} path has been elected as requiring authentication for all the tested router models.
    
    \item \textbf{Moxa Routers} present varying authentication schemas depending on the version of the server. In both cases, an HTML form is employed, being a password only form in \texttt{MoxaHttp1.0} routers, and a user and password form in \texttt{MoxaHttp2.0} routers. These routers, however, present the option of not having any password configured, giving direct access to the router web console in this case. For checking the validity of the entered credentials, as well as the lack of authentication, different text outputs provided as responses to the login are compared.
    \begin{itemize}
        \item \textbf{MoxaHttp1.0 routers} present a password only form in cases requiring authentication. Upon connection by the client, the router sends a random \texttt{FakeChallenge} field that is set on the client as a variable by the usage of JavaScript. At the moment of submitting the entered password, the client creates a MD5 hash of the entered password. This password, in conjunction with the \texttt{FakeChallenge} field, is then sent to the router via a GET request, as parameters to the \texttt{/home.htm} path. \\
    
        \begin{Verbatim}
            Password: cd7fb6d7e51b920b59ef1a6d310ac18e
            Submit: Submit
            token_text: 
            FakeChallenge: 28F87D79B5CF8608662E50F2ED7555CC4606E62DA647D103ED80007EB7D2320A
        \end{Verbatim}
        
        \item \textbf{MoxaHttp2.0 routers} present a user and password form in cases requiring authentication. Upon connection by the client, the router sends a random \texttt{FakeChallenge} field in form of a hidden input field in the HTML form. A second hidden HTML image input field captures the position of the login button.
        When submitting the password, on the client, the password hash is obtained first using the MD5 algorithm. This password hash is then sent in conjunction with the \texttt{FakeChallenge} field and the coordinates of the login button, in a POST request made to the base path.
        \begin{Verbatim}
            Username: admin
            Password: 
            MD5Password: 1cf153be978cb3
            FakeChallenge: 956901051
            Submit.x: 45
            Submit.y: 24
        \end{Verbatim}
        
    \end{itemize}
    
    \item \textbf{Sierra Wireless} routers present a HTML form in which the entered credentials are sent by an underlying XML request handler. When entering a credential, the client sends a XML encoded POST request to the \texttt{/xml/Connect.xml} path. This request includes both the username and password without any kind of encryption. Upon checking the validity of the credentials, the routers sends well defined XML responses with a message detailing if the login has been successful. This message is used as the method for checking the validity of the login in the scanner.
    \begin{Verbatim}[commandchars=\\\{\},codes={\catcode`$=3\catcode`_=8}]
        <request xmlns="urn:acemanager">
            <connect>
                <login>user</login>
                <password><![CDATA[password]]></password>
            </connect>
        </request>
    \end{Verbatim}
    
\end{itemize}

\subparagraph{Extensibility for other models} As happens for the various models of robots, extensibility has also been kept in mind when developing the \texttt{IndustrialRouterAdapter}. This scanner eases the inclusion of new industrial router models, by providing a base \texttt{IndustrialRouterScanner} class that can be extended for the various needs of each model. The scanners being used are then defined in the \texttt{router\_scanner\_types} class variable present in the \texttt{IndustrialRouterAdapter} class.\\

\noindent For the addition of the new model, the \texttt{IndustrialRouterScanner} class is extended, and the \texttt{possible\_headers}, \texttt{default\_credentials} and \texttt{router\_cls} variables defined.\\ The \texttt{possible\_headers} field defines the HTTP headers that identify the router as being of the defined model, while the \texttt{default\_credentials} and the \texttt{router\_cls} variables define the list of credentials to be tested, and the router class to be created as a result of a positive router find.\\

\noindent This base class provides a default method for testing credentials in HTTP basic authentication scenarios. In order to support other authentication schemes, the \texttt{check\_default\_password} is required to be extended.
The newly created class is then included in the mentioned \texttt{router\_scanner\_types} list in order to be used by the adapter.

\section{Results}
\label{scan_results}

\noindent For this study, different scans have been launched in search for robots and connected industrial routers. While in the case of ROS and SROS devices, direct, internet wide scans have been performed, in the case of the industrial routers, Shodan.io\cite{shodan} has been employed in order to cover the full internet spectrum. In the following sections, each of the scan methodology and the results are described.

\subsection{ROS and SROS robots}

\noindent During the research, two different global scans of the full internet address space have been performed. This scans have first searched for open ROS Master (11311) ports, and then \texttt{aztarna} has been used to check that these found hosts actually correspond to machines running ROS or SROS. Finally, the results provided by \texttt{aztarna} have followed a manual analysis to determine the nature of the found system and its location, among any other information that could be useful for the ongoing research. \\

\noindent For the initial scan for open ports, in an effort to improve the scan speed, the help of the ZMap tool has been required. ZMap is a tool optimized for massive scans, that has the ability send up to 14.2 million packets per second\cite{Adrian2014}, which in scans that only send one packet, such as the TCP SYN scan used in this research, translated to 14.2 million IP, addresses checked per second. This leads to performing a full internet scan in less than 5 minutes.\\

\noindent In the case of this study, the lack of dedicated hardware has leveraged to slower scan rates, which have caused the scan to span to a week.
To run the scan, 5 different virtual machines, running Ubuntu 18.04, each of them with 4 cores and 8Gb of RAM, deployed in well known cloud providers, have been used. The packet rate has been empirically chosen at 800pps, as higher rates showed packet drops that affected the scan effectiveness negatively.\\

\noindent To evenly distribute the work between the different nodes, the sharding capabilities that zmap provides have been used. These capabilities allow to divide the address space to search for each of the nodes, then converging the results.\\

\noindent After this first selection and target recollection process, \texttt{aztarna} has been used to determine the validity of the nodes found. For this, two scans have been launched with \texttt{aztarna}, one for ROS, in extended mode, and another one for SROS, in standard mode, as the complete host scan is time expensive to complete and could be interpreted as a phase preliminary to an attack by the target due to the port scan that is performed. In this case, a single machine has been used to perform both scans, as this scans are required to scan a lower number of possible targets. File output has been chosen to further analyze the data.\\

\noindent With the obtained data, manual analysis and filtering has been done, in order to determine the nature of the found robots, using the found topics and communications as indications to determine the model, purpose, and to determine if the instance is working as a simulated environment or not.\\

                                      

\begin{table}[H]
\caption{Scan Results for ROS systems by country.}
\centering
\begin{tabular}{lrrrrrrrr}
\cline{2-9}
                     & \multicolumn{4}{c|}{\cellcolor[HTML]{EFEFEF}\textbf{Scan 1}}                                                                & \multicolumn{4}{c|}{\cellcolor[HTML]{EFEFEF}\textbf{Scan 2}}                                                               \\ \hline
Country              & \multicolumn{1}{l}{Empty} & \multicolumn{1}{l}{Real} & \multicolumn{1}{l}{Simulation} & \multicolumn{1}{l|}{\textbf{Total}} & \multicolumn{1}{l}{Empty} & \multicolumn{1}{l}{Real} & \multicolumn{1}{l}{Simulation} & \multicolumn{1}{l}{\textbf{Total}} \\ \hline
AU                   & 0                         & 1                        & 0                              & \textbf{1}                          & 0                         & 1                        & 0                              & \textbf{1}                         \\
CA                   & 4                         & 1                        & 0                              & \textbf{5}                          & 0                         & 1                        & 1                              & \textbf{2}                         \\
CN                   & 2                         & 0                        & 0                              & \textbf{2}                          & 3                         & 2                        & 2                              & \textbf{7}                         \\
CZ                   & 0                         & 0                        & 0                              & \textbf{0}                          & 2                         & 0                        & 0                              & \textbf{2}                         \\
DE                   & 0                         & 0                        & 0                              & \textbf{0}                          & 1                         & 0                        & 1                              & \textbf{2}                         \\
ES                   & 1                         & 0                        & 0                              & \textbf{1}                          & 4                         & 0                        & 0                              & \textbf{4}                         \\
EU                   & 0                         & 0                        & 0                              & \textbf{0}                          & 0                         & 1                        & 0                              & \textbf{1}                         \\
GR                   & 0                         & 0                        & 0                              & \textbf{0}                          & 1                         & 4                        & 0                              & \textbf{5}                         \\
HK                   & 2                         & 2                        & 0                              & \textbf{4}                          & 2                         & 0                        & 0                              & \textbf{2}                         \\
IT                   & 1                         & 0                        & 0                              & \textbf{1}                          & 4                         & 1                        & 0                              & \textbf{5}                         \\
JP                   & 2                         & 0                        & 0                              & \textbf{2}                          & 1                         & 0                        & 0                              & \textbf{1}                         \\
KR                   & 5                         & 0                        & 3                              & \textbf{8}                          & 6                         & 4                        & 6                              & \textbf{16}                        \\
NL                   & 1                         & 0                        & 0                              & \textbf{1}                          & 1                         & 0                        & 0                              & \textbf{1}                         \\
SE                   & 1                         & 0                        & 0                              & \textbf{1}                          & 0                         & 0                        & 0                              & \textbf{0}                         \\
SG                   & 0                         & 0                        & 0                              & \textbf{0}                          & 1                         & 0                        & 0                              & \textbf{1}                         \\
TW                   & 2                         & 0                        & 0                              & \textbf{2}                          & 2                         & 0                        & 2                              & \textbf{4}                         \\
US                   & 21                        & 7                        & 0                              & \textbf{28}                         & 25                        & 22                       & 5                              & \textbf{52}                        \\ \hline
\textbf{Grand Total} & \textbf{42}               & \textbf{11}              & \textbf{3}                     & \textbf{56}                         & \textbf{53}               & \textbf{36}              & \textbf{17}                    & \textbf{106}                       \\ \hline
\end{tabular}
\label{table:ros_results}
\end{table}

\begin{figure}[H]
\centering
 \includegraphics[width=0.7\textwidth]{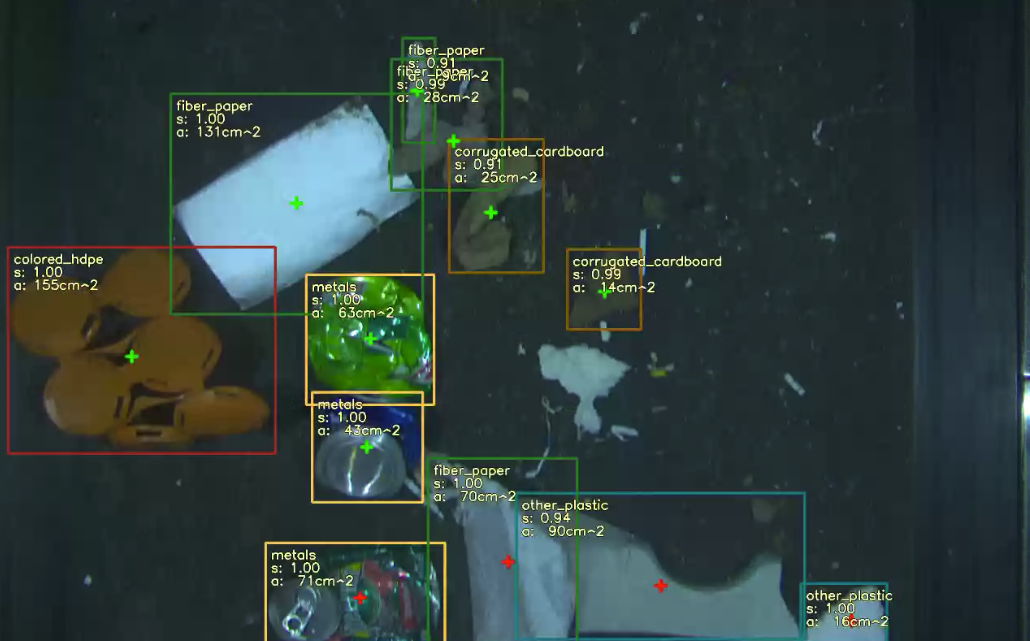}
 \caption{\footnotesize Example of industrial, trash classification robot, found with \texttt{aztarna}}
 \label{fig:classification_robot}
\end{figure}

\subsection{Industrial Routers}

\noindent For industrial routers, a single internet wide scan has been launched, using Shodan in order to determine the test targets by the defined values to query. This scan has revealed a vast amount of connected devices, many of them using default, well known credentials, or having no authentication mechanisms at all. For this scan, routers from Ewon, Moxa, Westermo and Sierra Wireless manufactures have been taken into account. This manufacturers represent the majority of industrial routers present in the industry. 
As it can be seen on table \ref{table:industrial_routers_by_manufacturer}, from the returned 61265 results from Shodan, 26801 have been determined as answering to connections, and from this routers, a total of 8958 routers have been determined to be configured with default credentials or no authentication, which represents a 33\% of the analyzed routers.\\

\noindent The difference between the results returned by Shodan and by the then alive routers is presumably caused by the usage of dynamic IP addressing in the connections utilized by the routers. As mentioned before, Shodan utilizes a variety of crawlers that periodically scan the internet searching for connected devices. The time difference between the scan by Shodan and the scan by \texttt{aztarna} may affect the detection of the devices. On an empirical basis, some of the devices that were being individually analyzed stopped answering in a time-span of 8 hours, showing a change in the assigned IP address.\\

\begin{figure}[H]
\centering
 \includegraphics[width=0.7\textwidth]{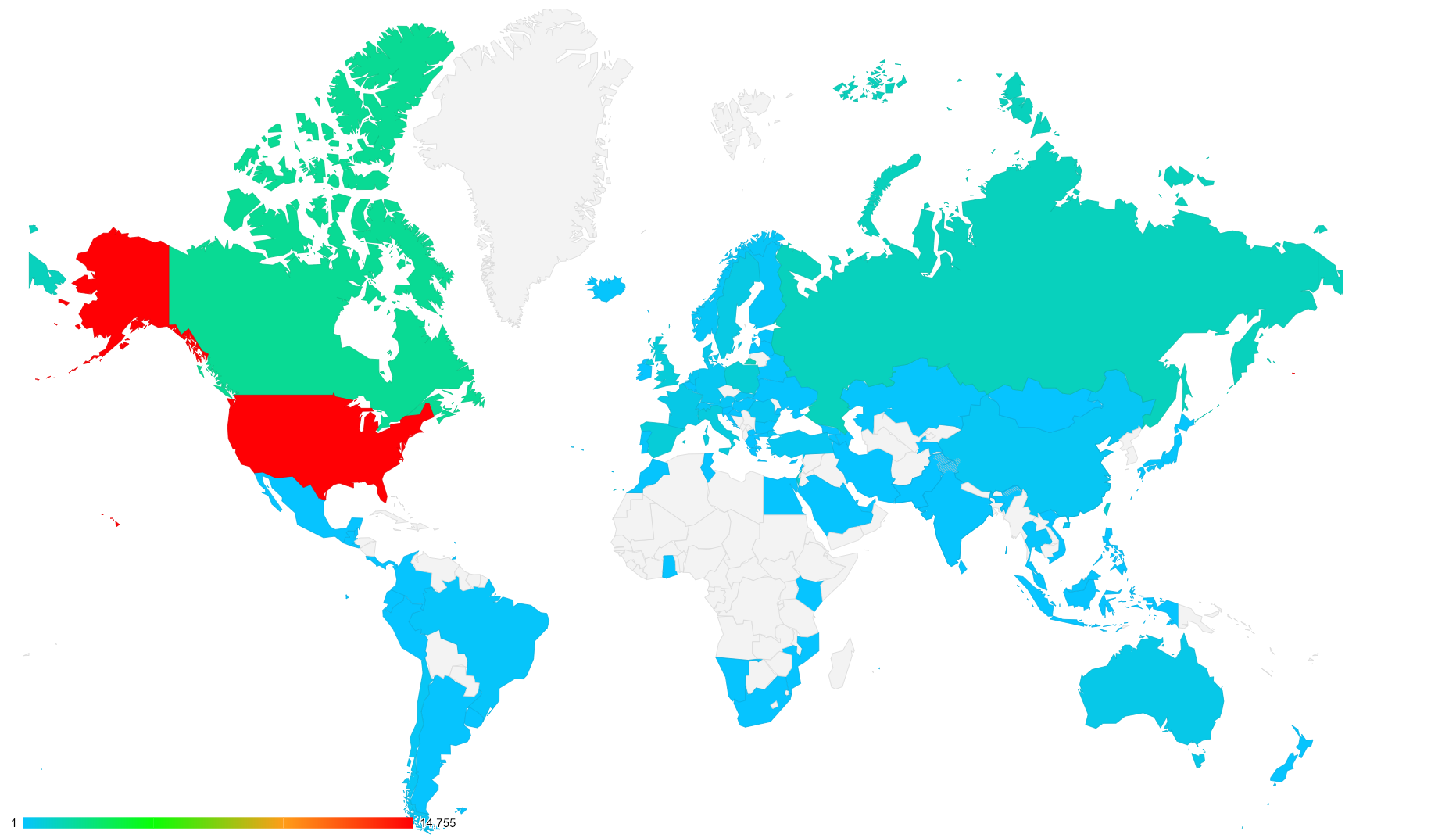}
 \caption{\footnotesize Industrial router quantity by country}
 \label{fig:router_quantity}
\end{figure}

\begin{figure}[H]
\centering
 \includegraphics[width=0.7\textwidth]{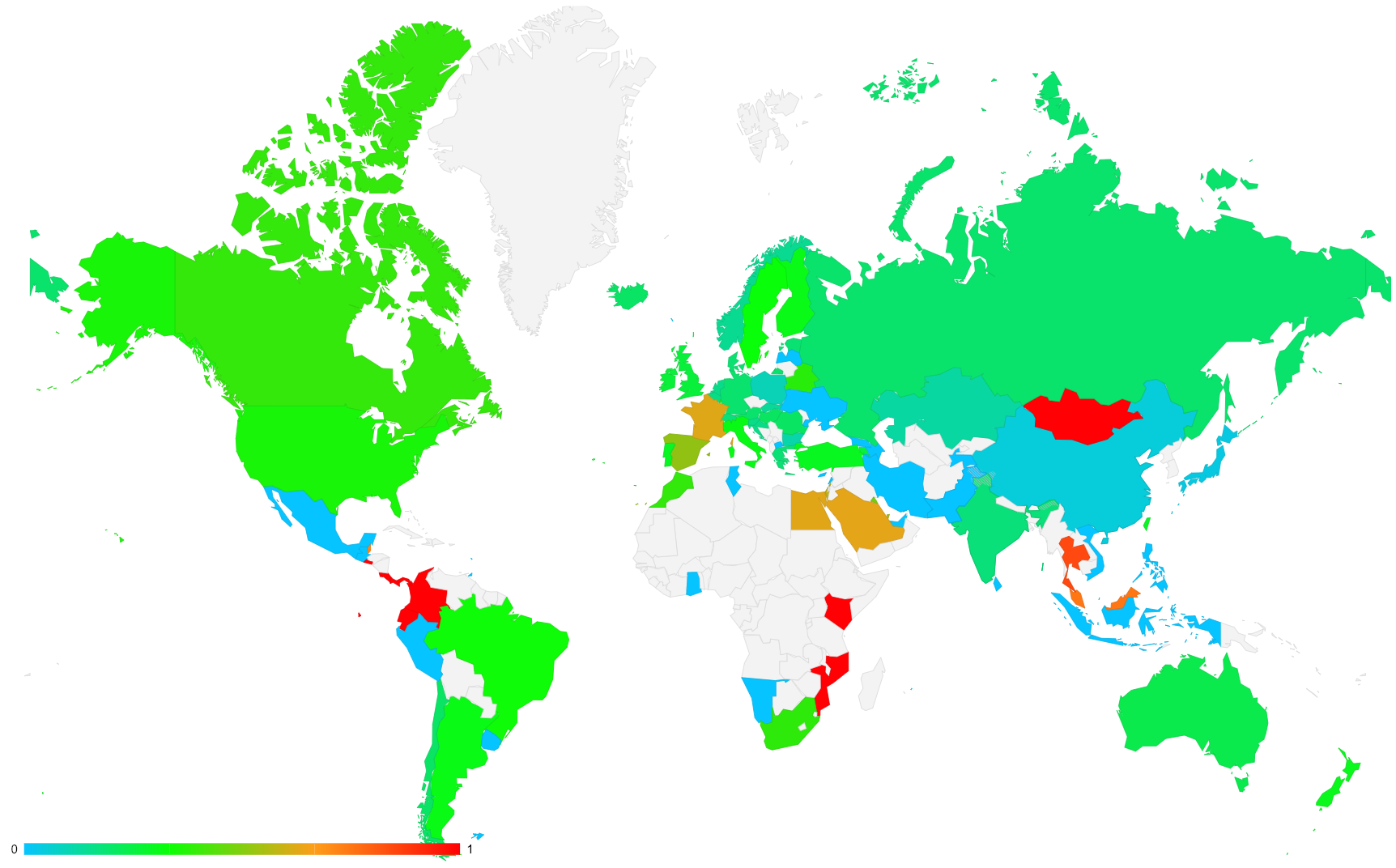}
 \caption{\footnotesize Proportion of industrial routers using no authentication or default credentials. From 0 (0\%, blue) of the devices using default credentials, to 1 (100\%, red) of the devices using default credentials }
 \label{fig:router_proportion}
\end{figure}

\noindent The resulting router quantity, as well as the proportion of the routers configured with default credentials, vs the ones with changed credentials is detailed on table \ref{table:routers_by_country}, as well as in the included proportion \ref{fig:router_proportion} and quantity \ref{fig:router_quantity} heat maps by country.
These results show that United States is the country showing the highest number of connected devices (14755), followed by Canada (1869) and Russia(1120). 
Most countries follow a similar balance between correctly configured and misconfigured devices, being Colombia, with 26 connected devices and with 100\% of the devices using default credentials, the most insecure country. Thailand follows the ranking with 54 devices, showing 93\% of them using default credentials. 
From the countries with a higher number of connected devices, France stands out in the proportion of misconfigured devices, having a total of 416 devices, with 261 of them(63\%) configured with default credentials.

\begin{table}[H]
\centering
\caption{Scan results for industrial routers by manufacturer}
\begin{tabular}{lrrr}
\hline
\rowcolor[HTML]{EFEFEF} 
\multicolumn{1}{c}{\cellcolor[HTML]{EFEFEF}\textbf{Type}} & \multicolumn{1}{c}{\cellcolor[HTML]{EFEFEF}\textbf{Detected}} & \multicolumn{1}{c}{\cellcolor[HTML]{EFEFEF}\textbf{Alive}} & \multicolumn{1}{c}{\cellcolor[HTML]{EFEFEF}\textbf{Default password}} \\ \hline
Ewon                                                      & 4465                                                          & 983                                                        & 359                                                                   \\
Moxa                                                      & 13291                                                         & 6549                                                       & 971                                                                   \\
Westermo                                                  & 4602                                                          & 1393                                                       & 279                                                                   \\
Sierra Wireless                                           & 38907                                                         & 17876                                                      & 7400                                                                  \\ \hline
\textbf{Grand Total}                                            & \textbf{61265}                                                & \textbf{26801}                                             & \textbf{9009}                                                         \\ \hline
\end{tabular}
\label{table:industrial_routers_by_manufacturer}
\end{table}

\begin{longtable}{lrrrr}
\caption{Scan results for industrial routers by country}

\\ \hline
\multicolumn{1}{c}{\cellcolor[HTML]{EFEFEF}\textbf{Country}} & \multicolumn{1}{c}{\cellcolor[HTML]{EFEFEF}\textbf{Routers}} & \multicolumn{1}{c}{\cellcolor[HTML]{EFEFEF}\textbf{Default Credentials}} & \multicolumn{1}{c}{\cellcolor[HTML]{EFEFEF}\textbf{Changed credentials}} & \multicolumn{1}{c}{\cellcolor[HTML]{EFEFEF}\textbf{Proportion}} \\ \hline \\
\endhead
US     & 14755 & 5383 & 9372 & 36\%   \\
CA     & 1869  & 767  & 1102 & 41\%   \\
RU     & 1120  & 201  & 919  & 18\%   \\
PL     & 706   & 59   & 647  & 8\%    \\
TW     & 702   & 197  & 505  & 28\%   \\
ES     & 664   & 356  & 308  & 54\%   \\
IL     & 660   & 393  & 267  & 60\%   \\
GB     & 639   & 158  & 481  & 25\%   \\
IT     & 556   & 169  & 387  & 30\%   \\
SE     & 453   & 149  & 304  & 33\%   \\
FR     & 416   & 261  & 155  & 63\%   \\
AU     & 395   & 87   & 308  & 22\%   \\
DK     & 333   & 57   & 276  & 17\%   \\
DE     & 238   & 42   & 196  & 18\%   \\
NL     & 219   & 31   & 188  & 14\%   \\
TR     & 209   & 60   & 149  & 29\%   \\
RO     & 204   & 39   & 165  & 19\%   \\
CN     & 190   & 8    & 182  & 4\%    \\
AT     & 174   & 31   & 143  & 18\%   \\
CL     & 171   & 32   & 139  & 19\%   \\
SK     & 165   & 20   & 145  & 12\%   \\
CH     & 162   & 26   & 136  & 16\%   \\
NO     & 131   & 17   & 114  & 13\%   \\
FI     & 118   & 36   & 82   & 31\%   \\
PT     & 96    & 27   & 69   & 28\%   \\
KR     & 88    & 57   & 31   & 65\%   \\
CZ     & 84    & 10   & 74   & 12\%   \\
HK     & 83    & 10   & 73   & 12\%   \\
BR     & 73    & 25   & 48   & 34\%   \\
EE     & 64    & 11   & 53   & 17\%   \\
IS     & 64    & 12   & 52   & 19\%   \\
LT     & 60    & 9    & 51   & 15\%   \\
JP     & 58    & 2    & 56   & 3\%    \\
KZ     & 55    & 6    & 49   & 11\%   \\
TH     & 54    & 45   & 9    & 83\%   \\
HU     & 52    & 10   & 42   & 19\%   \\
BG     & 51    & 5    & 46   & 10\%   \\
MY     & 46    & 34   & 12   & 74\%   \\
BY     & 41    & 16   & 25   & 39\%   \\
BE     & 34    & 5    & 29   & 15\%   \\
AM     & 33    & 8    & 25   & 24\%   \\
MA     & 32    & 13   & 19   & 41\%   \\
IN     & 31    & 5    & 26   & 16\%   \\
UA     & 30    & 0    & 30   & 0\%    \\
LV     & 30    & 0    & 30   & 0\%    \\
SG     & 29    & 2    & 27   & 7\%    \\
Others & 393   & 120   & 273   & 30\% \\
\hline
\textbf{Grand Total} & \textbf{26801} & \textbf{9009} & \textbf{17792} & \textbf{34\%}                                                       \\ \hline

\end{longtable}
\label{table:routers_by_country}

\section{Remarks and future work}
\label{conclusion}

\noindent ROS was born as a research framework for robotics development. Now, it's being gradually replaced by its  second version, ROS 2. This new version has taken a massive architectural shift, that will require a separate reconnaissance rationale. ROS 2 technology is based on the DDS (Data Distribution Service) standard, which allows ROS 2 based systems to communicate in a low latency, extremely reliable and distributed environment\cite{dds}. ROS 2 security is currently under discussion.\\

\noindent Apart from ROS, some robot manufacturers tend to develop their own exclusive programming APIs. Although ROS is becoming the \emph{de facto} standard in robot programming, there is a set of proprietary tools being extensively used by leading robotic companies. One example is ABB's RobAPI \cite{robapi}, a REST based programming library.\\

\noindent Extensions of \texttt{aztarna} towards \emph{fingerprinting} are also expected. Examples of such extensions include determining the specific firmware version in robots, discovering third-party libraries used and their versions (e.g. robot middleware version, communication infrastructure, etc.). In industrial environments, robots commonly use industrial communication protocols. These can be both standardized (e.g. EtherCAT, PROFINET) or proprietary (e.g. ABB's MMS, Beckhoffs's ADS/AMS). The ability to dissect and understand these protocols is itself useful for footprinting purposes.\\

\noindent With regards to unprotected industrial routers in this work, there are other models that could be detected in a similar fashion as the ones presented in subsection \ref{subsec:irouters}, just by looking at their HTTP response headers, but are not included within the present version of \texttt{aztarna}. Other techniques to locate vulnerable industrial routers include the FTP port scanning and banner grabbing, which is demonstrated in \cite{Maggi} are not explored in the present version. These include \textit{Sierra Wireless} and \textit{Digi} among others \cite{Maggi}, widely used in industrial environments.\\

\noindent Industrial routers pose an entry point for robots that are located on the network behind the industrial router, which is usually directly connected to the device. This is more evident in the cases of routers exposing default passwords. Most of these routers have reported vulnerabilities and flaws that are subject to update by users. Furthermore, given that most industrial routers provide means to establishing VPN connections, both in server and client mode, tunnels to a protected network behind the routers are a very likely attack vector The automation of this processes could allow researchers to audit the devices that are behind of the router, as well as the router security itself, providing a better insight on the nature of the target. This process will be part of the industrial routers fingerprinting phase, as we pretend to go deep exploring new vulnerabilities.\\

\noindent With the retrieved information, we aim to reinforce the awareness of the robot users community we advocate in favour of secure and tested industrial networks to avoid the dangers of targeted attacks towards robots.  Finally, we encourage the manufacturers to address a security by design policy. \\

\noindent Overall, we conclude that \texttt{aztarna} responds to the need of auditing robot security. As ROS is becoming the \emph{de facto} standard in robot programming, more and more robots are being exposed everyday. The footprinting techniques on ROS are specially dangerous, because once detected and footprinted, ROS powered systems are inherently vulnerable. Existing robot security mitigations, such as SROS, are not used extensively. The present study reports mainly research robots aligned with prior art, but we have reported the footprinting of professional robots as well. We have discovered an array of internet-connected unprotected industrial routers, that could potentially host robots. There is an unresolved gap in robotics cybersecurity which would greatly benefit from releasing the first auditing tools.\\

\section{Acknowledgments}
\noindent This research has been partially funded by the Basque Government, throughout the Business Development Agency of the Basque Country (SPRI) through the \emph{Ekintzaile} 2018 program and EU H2020 Robot Union Program through the  Grant Agreement nº 779967. Special thanks to BIC Araba and the Basque Cybersecurity Centre (BCSC) for the support provided. 

\appendix

\bibliographystyle{IEEEtran}
\bibliography{bibliography}
\end{document}